\documentclass[final,3p,times]{elsarticle}
\usepackage{amssymb}
\usepackage[fleqn]{amsmath}
\usepackage{amsthm}
\usepackage{subfigure}
\usepackage{graphicx}

\biboptions{numbers,sort&compress}
\usepackage{hyperref}

\begin{document}

\begin{frontmatter}



\title{Rogue wave solutions in AB system }

\author{Xin Wang} \author{Yuqi Li}
\author{Yong Chen\corref{cor1}}
\ead{ychen@sei.ecnu.edu.cn}

\cortext[cor1]{Corresponding author.}

\address{Shanghai Key Laboratory of Trustworthy Computing, East China Normal University, Shanghai, 200062, People's Republic of China}

\begin{abstract}
In this paper, the generalized Darboux transformation is established to the AB system, which mainly
describes marginally unstable baroclinic wave packets in geophysical fluids and ultra-short pulses in nonlinear optics.
We find a unified formula of $N$th-order rogue wave solution for the AB system by the direct iterative rule. In particular, rogue waves
possessing  several free parameters from first to second order are calculated.  The dynamics and some interesting structures of the rogue waves are
illustrated through some figures.
\end{abstract}

\begin{keyword}
Generalized Darboux transformation;  Rogue wave; AB system
\end{keyword}
\end{frontmatter}



\section{Introduction}  
It is well known that the AB system serves as model equations
to describe marginally unstable baroclinic wave packets in geophysical fluids \cite{1}, ultra-short optical pulse propagation in nonlinear optics \cite{2}, and also mesoscale gravity current transmission on a sloping bottom in the problem of cold gravity current \cite{3}. It was firstly proposed by Pedlosky using the singular perturbation theory \cite{1}. So far, there has been surge of interesting in studying the dynamics
properties of the AB system, such as the single-phase periodic solution depending on a complete set of four complex parameters \cite{4}, the envelope solitary waves and periodic waves \cite{5}, the Painlev\'{e} analysis and conservation laws of the variable-coefficient AB system \cite{6}, the soliton and  breather solutions through the classical Darboux transformation \cite{7}, and the N-soliton solutions by using the dressing method \cite{8}.

To our knowledge, there are no reports on rogue
waves of the AB system up to the present. While
in the past five years, rogue waves (also known as freak waves, monster waves, killer waves, rabid-dog waves
and similar names) \cite{9,10,11,12}
have become a hot spot in the research field of oceanography \cite{14}, optics \cite{15}, Bose-Einstein condensates \cite{16},
superfluid \cite{17}, capillary flow \cite{18} and even finance \cite{19}. A rogue wave  is localized in both space and time, and can be depicted as a wave which appears from nowhere and disappears without a trace \cite{13}.
Many nonlinear Schr\"{o}dinger (NLS)-type equations, for instance, the standard NLS equation \cite{11,22,23,35,36},
the Hirota equation \cite{24}, the Sasa-Satsuma equation \cite{25}, the high-order dispersive generalized NLS equation \cite{26},
the variable coefficient  NLS equation \cite{27}, the discrete NLS equation \cite{28},
the Manakov equations \cite{29,30,31}, the coupled Hirota equations \cite{32}, the three-component NLS equations \cite{37}
have been confirmed to possess lower or high-order rogue waves of diverse  structures. Nevertheless,
there are relatively few papers on rogue waves of the non-NLS-type equations, and a complete understanding of the mysterious and
catastrophic rogue wave phenomenon is still far from been achieved, due to the difficult and hazardous observational conditions \cite{13}.
Therefore, it is of great interest
to investigate  rogue waves of the non-NLS-type AB system, which may be helpful to better understand the dynamics
properties of the complicated rogue wave phenomenon.

In this paper, we take the  AB system  in  canonical form \cite{4}
\begin{align}
&A_{xt}=AB, \label{01}\\
&B_{x}=-\frac{1}{2}(|A|^{2})_{t},\label{02}
\end{align}
where  $x$ and $t$ are semi-characteristic normalized coordinates, $A$ and $B$ are the wave amplitudes yielding the normalization condition
\begin{equation}\label{03}
|A_{t}|^{2}+B^{2}=1.
\end{equation}
When $A$ is the real value, Eqs. (1) and (2) can be transformed into the Sine-Gordon equation, and
when A is the complex value, the self-induced transparency system \cite{2,4}.

The aim of the present paper is to research Eqs. (1) and (2) through the so-called generalized Darboux transformation (DT) \cite{23}, which
been proposed by Guo, Ling and Liu, has provided a powerful tool to derive general rogue wave solutions of many
nonlinear equations, the NLS equation \cite{23}, the derivative NLS equation \cite{33}, the Manakov equations \cite{34}, etc.
With the help of the generalized DT, a unified formula of $N$th-order rogue wave solution for Eqs. (1) and (2)
is obtained by the direct iterative rule. As application, rogue waves of Eqs. (1) and (2) from first to second order are
studied.
The first-order rogue waves of fundamental pattern, the second-order rogue waves of
fundamental and triangular patterns are displayed by  choosing different parameters,  respectively.

Our paper is organized as follows. In section 2, we construct the  generalized DT to Eqs. (1) and (2) under the normalization condition (\ref{03}),
then a unified formula of $N$th-order rogue wave solution for Eqs. (1) and (2) is obtained by the direct iterative rule.
In section 3, the dynamics and some interesting structures of the rogue waves for Eqs. (1) and (2) from first to second order are
illustrated through some figures. In section 4, we give the conclusion.

\section{Generalized Darboux transformation}
In this section, we start from the Lax pair of Eqs. (1) and (2), which reads
\begin{align}
&\Psi_{x}=U\Psi,\ U=\left(
\begin{array}{cc}
-{\rm i}\lambda & \frac{1}{2}A \\
-\frac{1}{2}A^{*} & {\rm i}\lambda \\
\end{array}
\right), \label{04}\\
&\Psi_{t}=V\Psi,\ V=\frac{1}{4{\rm i}\lambda}
\left(
  \begin{array}{cc}
    -B & A_{t} \\
    A^{*}_{t} & B \\
  \end{array}
\right),\label{05}
\end{align}
where $\Psi=(\psi(x,t),\phi(x,t))$ is the vector eigenfunction, $\lambda$ is the spectral parameter, and asterisk denotes
the complex conjugation. It could be easily verified that the compatibility condition $U_{t}-V_{x}+UV-VU=0$ gives rise to Eqs. (1) and (2).

Next, let  $\Psi_{1}=(\psi_{1},\phi_{1})$ be a basic solution of the Lax pair (\ref{04}) and (\ref{05}) with $A=A[0]$, $B=B[0]$ and $\lambda=\lambda_{1}$. Thus,
on basis of the above lax pair, the classical DT \cite{20,21} of Eqs. (1) and (2) can be built \cite{7}
\begin{align}
&\Psi[1]=T[1]\Psi,\ T[1]=(\lambda I-H[0]\Lambda_{1}H[0]^{-1}),  \label{06}\\
&A[1]=A[0]-4{\rm i}(\lambda_{1}-\lambda_{1}^{*})\frac{\psi_{1}[0]\phi_{1}[0]^{*}}{(|\psi_{1}[0]|^{2}+|\phi_{1}[0]|^{2})},\label{07}\\
&B[1]=B[0]-4{\rm i}(\lambda_{1}-\lambda_{1}^{*})\frac{[|\psi_{1}[0]|^{2}(|\phi_{1}[0]|^{2})_{t}-|\phi_{1}[0]|^{2}(|\psi_{1}[0]|^{2})_{t}]}{(|\psi_{1}[0]|^{2}
+|\phi_{1}[0]|^{2})^{2}},\label{08}
\end{align}
where $\psi_{1}[0]=\psi_{1}$, $\phi_{1}[0]=\phi_{1}$,
$$
I=\left(
    \begin{array}{cc}
      1 & 0 \\
      0 & 1 \\
    \end{array}
  \right),\
H[0]=
\left(
  \begin{array}{cc}
    \psi_{1}[0] & \phi_{1}[0]^{*} \\
    \phi_{1}[0] & -\psi_{1}[0]^{*} \\
  \end{array}
\right),\ \Lambda_{1}=
\left(
  \begin{array}{cc}
    \lambda_{1} & 0 \\
    0 & \lambda_{1}^{*} \\
  \end{array}
\right).
$$
In the following, suppose $\Psi_{l}=(\psi_{l},\phi_{l}),~1\leq l\leq N$ be a basic solution of the Lax pair (\ref{04}) and (\ref{05}) with
$A=A[0]$, $B=B[0]$ and $\lambda=\lambda_{l}$. Then the $N$-step classical DT of Eqs. (1) and (2) can be  naturally given  as follows
\begin{align}
&\Psi[N]=T[N]T[N-1]\cdots T[1]\Psi, \ T[l]=\lambda I-H[l-1]\Lambda_{l}H[l-1]^{-1}, \label{09}\\
&A[N]=A[N-1]-4{\rm i}(\lambda_{N}-\lambda_{N}^{*})\frac{\psi_{N}[N-1]\phi_{N}[N-1]^{*}}{(|\psi_{N}[N-1]|^{2}+|\phi_{N}[N-1]|^{2})},\label{10}\\
&B[N]=B[N-1]-4{\rm i}(\lambda_{N}-\lambda_{N}^{*})\frac{[|\psi_{N}[N-1]|^{2}(|\phi_{N}[N-1]|^{2})_{t}-|\phi_{N}[N-1]|^{2}(|\psi_{N}[N-1]|^{2})_{t}]}{(|\psi_{N}[N-1]|^{2}
+|\phi_{N}[N-1]|^{2})^{2}},\label{11}
\end{align}
where
$$
H[l-1]=
\left(
  \begin{array}{cc}
    \psi_{l}[l-1] & \phi_{l}[l-1]^{*} \\
    \phi_{l}[l-1] & -\psi_{l}[l-1]^{*} \\
  \end{array}
\right),\ \Lambda_{l}=
\left(
  \begin{array}{cc}
    \lambda_{l} & 0 \\
    0 & \lambda_{l}^{*} \\
  \end{array}
\right),
$$
with $(\psi_{l}[l-1],\phi_{l}[l-1])=\Psi_{l}[l-1]$, and
$$
\Psi_{l}[l-1]=T_{l}[l-1]T_{l}[l-2]\cdots T_{l}[1]\Psi_{l},\  T_{l}[k]=T[k]|_{\lambda=\lambda_{l}},\ 1\leq l\leq N,~1\leq k\leq l-1.
$$

According to the above facts, the generalized DT can be derived for Eqs. (1) and (2).  To this end, let $\Psi_{1}(\lambda_{1}+\delta)$ be  a special
solution of the Lax pair (\ref{04}) and (\ref{05}) with $A[0]$, $B[0]$ and $\lambda=\lambda_{1}+\delta$, and it can be expanded as Taylor series
at $\delta=0$, that is
\begin{equation}\label{12}
\Psi_{1}=\Psi_{1}^{[0]}+\Psi_{1}^{[1]}\delta+\Psi_{1}^{[2]}\delta^{2}+\Psi_{1}^{[3]}\delta^{3}+\cdots+\Psi_{1}^{[N]}\delta^{N}+o(\delta^{N}),
\end{equation}
where $\Psi_{1}^{[k]}=(\psi_{1}^{[k]},\phi_{1}^{[k]})=\lim\limits_{\delta\rightarrow0}\frac{\displaystyle1}{\displaystyle k!}
\frac{\displaystyle\partial^{k}\Psi_{1}}{\displaystyle\partial \delta^{k}},~k=0,1,2,\cdots.$

Afterwards, it is easy to find that $\Psi_{1}^{[0]}$ is a special solution of the Lax pair (\ref{04}) and (\ref{05}) with $A=A[0]$, $B=B[0]$ and $\lambda=\lambda_{1}$.
Hence, by means of the formulas (\ref{06})-(\ref{08}), the first-step generalized DT of Eqs. (1) and (2) can be directly obtained.

(1) The first-step generalized DT
\begin{align}
&\Psi[1]=T[1]\Psi,\ T[1]=(\lambda I-H[0]\Lambda_{1}H[0]^{-1}),  \label{13}\\
&A[1]=A[0]-4{\rm i}(\lambda_{1}-\lambda_{1}^{*})\frac{\psi_{1}[0]\phi_{1}[0]^{*}}{(|\psi_{1}[0]|^{2}+|\phi_{1}[0]|^{2})},\label{14}\\
&B[1]=B[0]-4{\rm i}(\lambda_{1}-\lambda_{1}^{*})\frac{[|\psi_{1}[0]|^{2}(|\phi_{1}[0]|^{2})_{t}-|\phi_{1}[0]|^{2}(|\psi_{1}[0]|^{2})_{t}]}{(|\psi_{1}[0]|^{2}
+|\phi_{1}[0]|^{2})^{2}},\label{15}
\end{align}
where $\psi_{1}[0]=\psi_{1}^{[0]}$, $\phi_{1}[0]=\phi_{1}^{[0]}$,
$$
H[0]=
\left(
  \begin{array}{cc}
    \psi_{1}[0] & \phi_{1}[0]^{*} \\
    \phi_{1}[0] & -\psi_{1}[0]^{*} \\
  \end{array}
\right),\ \Lambda_{1}=
\left(
  \begin{array}{cc}
    \lambda_{1} & 0 \\
    0 & \lambda_{1}^{*} \\
  \end{array}
\right).
$$

(2) The second-step generalized DT

It is clear that $T[1]\Psi_{1}$ is a basic solution of the Lax pair (\ref{04}) and (\ref{05}) with $A[1]$, $B[1]$ and $\lambda=\lambda_{1}+\delta$.
So that, by using  the identity $T_{1}[1]\Psi_{1}^{[0]}=0$, the following limit process
$$
\lim\limits_{\delta\rightarrow0}\frac{\displaystyle T[1]|_{\lambda=\lambda_{1}+\delta}\Psi_{1}}{\displaystyle\delta}=\lim\limits_{\delta\rightarrow0}
\frac{\displaystyle(\delta+T_{1}[1])\Psi_{1}}{\displaystyle\delta}=
\Psi_{1}^{[0]}+T_{1}[1]\Psi_{1}^{[1]}\equiv\Psi_{1}[1]
$$
provides a nontrivial solution of the Lax pair (\ref{04}) and (\ref{05}) with $A[1]$, $B[1]$, $\lambda=\lambda_{1}$, and can be adopted
to do the second-step generalized DT, i.e.
\begin{align}
&\Psi[2]=T[2]T[1]\Psi,\ T[2]=(\lambda I-H[1]\Lambda_{2}H[1]^{-1}),  \label{16}\\
&A[2]=A[1]-4{\rm i}(\lambda_{1}-\lambda_{1}^{*})\frac{\psi_{1}[1]\phi_{1}[1]^{*}}{(|\psi_{1}[1]|^{2}+|\phi_{1}[1]|^{2})},\label{17}\\
&B[2]=B[1]-4{\rm i}(\lambda_{1}-\lambda_{1}^{*})\frac{[|\psi_{1}[1]|^{2}(|\phi_{1}[1]|^{2})_{t}-|\phi_{1}[1]|^{2}(|\psi_{1}[1]|^{2})_{t}]}{(|\psi_{1}[1]|^{2}
+|\phi_{1}[1]|^{2})^{2}},\label{18}
\end{align}
where $(\psi_{1}[1],\phi_{1}[1])^{T}=\Psi_{1}[1]$,
$$
H[1]=
\left(
  \begin{array}{cc}
    \psi_{1}[1] & \phi_{1}[1]^{*} \\
    \phi_{1}[1] & -\psi_{1}[1]^{*} \\
  \end{array}
\right),\ \Lambda_{2}=
\left(
  \begin{array}{cc}
    \lambda_{1} & 0 \\
    0 & \lambda_{1}^{*} \\
  \end{array}
\right).
$$

(3) The third-step generalized DT

In the same way, with the aid of the identities
$$T_{1}[1]\Psi_{1}^{[0]}=0,\ T_{1}[2](\Psi_{1}^{[0]}+T_{1}[1]\Psi_{1}^{[1]})=0,$$
we get the following limit process
$$\lim\limits_{\delta\rightarrow0}\frac{\displaystyle [T[2]T[1]]|_{\lambda=\lambda_{1}+\delta}\Psi_{1}}{\displaystyle\delta^{2}}=
\lim\limits_{\delta\rightarrow0}\frac{\displaystyle (\delta+T_{1}[2])(\delta+T_{1}[1])\Psi_{1}}{\displaystyle\delta^{2}}=
\Psi_{1}^{[0]}+(T_{1}[2]+T_{1}[1])\Psi_{1}^{[1]}+T_{1}[2]T_{1}[1]\Psi_{1}^{[2]}\equiv\Psi_{1}[2],
$$
which is a nontrivial solution of the Lax pair (\ref{04}) and (\ref{05}) with $A[2]$, $B[2]$, $\lambda=\lambda_{1}$, and can lead to
the third-step generalized DT, namely,
\begin{align}
&\Psi[3]=T[3]T[2]T[1]\Psi,\ T[3]=(\lambda I-H[2]\Lambda_{3}H[2]^{-1}),  \label{19}\\
&A[3]=A[2]-4{\rm i}(\lambda_{1}-\lambda_{1}^{*})\frac{\psi_{1}[2]\phi_{1}[2]^{*}}{(|\psi_{1}[2]|^{2}+|\phi_{1}[2]|^{2})},\label{20}\\
&B[3]=B[2]-4{\rm i}(\lambda_{1}-\lambda_{1}^{*})\frac{[|\psi_{1}[2]|^{2}(|\phi_{1}[2]|^{2})_{t}-|\phi_{1}[2]|^{2}(|\psi_{1}[2]|^{2})_{t}]}{(|\psi_{1}[2]|^{2}
+|\phi_{1}[2]|^{2})^{2}},\label{21}
\end{align}
where $(\psi_{1}[2],\phi_{1}[2])^{T}=\Psi_{1}[2]$,
$$
H[2]=
\left(
  \begin{array}{cc}
    \psi_{1}[2] & \phi_{1}[2]^{*} \\
    \phi_{1}[2] & -\psi_{1}[2]^{*} \\
  \end{array}
\right),\ \Lambda_{3}=
\left(
  \begin{array}{cc}
    \lambda_{1} & 0 \\
    0 & \lambda_{1}^{*} \\
  \end{array}
\right).
$$

(4) The $N$-step generalized DT.

Iterating the above process $N$ times, we ge the $N$-step generalized DT of Eqs. (1) and (2)
$$
\Psi_{1}[N-1]=\Psi_{1}^{[0]}+\sum_{l=1}^{N-1}T_{1}[l]\Psi_{1}^{[1]}+\sum_{l=1}^{N-1}\sum_{k=1}^{l-1}T_{1}[l]T_{1}[k]\Psi_{1}^{[2]}+\cdots+T_{1}[N-1]T_{1}[N-2]
\cdots T_{1}[1]\Psi_{1}^{[N-1]},
$$
\begin{align}
&\Psi[N]=T[N]T[N-1]\cdots T[1]\Psi,\ T[N]=(\lambda I-H[N-1]\Lambda_{N}H[N-1]^{-1}),  \label{22}\\
&A[N]=A[N-1]-4{\rm i}(\lambda_{1}-\lambda_{1}^{*})\frac{\psi_{1}[N-1]\phi_{1}[N-1]^{*}}{(|\psi_{1}[N-1]|^{2}+|\phi_{1}[N-1]|^{2})},\label{23}\\
&B[N]=B[N-1]-4{\rm i}(\lambda_{1}-\lambda_{1}^{*})\frac{[|\psi_{1}[N-1]|^{2}(|\phi_{1}[N-1]|^{2})_{t}-|\phi_{1}[N-1]|^{2}(|\psi_{1}[N-1]|^{2})_{t}]}{(|\psi_{1}[N-1]|^{2}
+|\phi_{1}[N-1]|^{2})^{2}},\label{24}
\end{align}
where $(\psi_{1}[N-1],\phi_{1}[N-1])^{T}=\Psi_{1}[N-1]$,
$$
H[l-1]=
\left(
  \begin{array}{cc}
    \psi_{1}[l-1] & \phi_{1}[l-1]^{*} \\
    \phi_{1}[l-1] & -\psi_{1}[l-1]^{*} \\
  \end{array}
\right),\ \Lambda_{l}=
\left(
  \begin{array}{cc}
    \lambda_{1} & 0 \\
    0 & \lambda_{1}^{*} \\
  \end{array}
\right),\ 1\leq l\leq N.
$$
By making use of the above formulas  (\ref{22})-(\ref{24}), a unified formula of $N$th-order rogue wave solution for Eqs. (1) and (2) can be obtained
by the direct iterative rule. Moreover, it is not difficult to convert (\ref{22})-(\ref{24}) into the $2N\times 2N$ determinant representation.
But, to avoid the calculation of the determinant of a matrix of very high order, we prefer to use Darboux transformations of degree one successively instead of
a  Darboux transformation of higher degree with  determinant representation. In the next section, the formulas (\ref{22})-(\ref{24}) will be
applied to work out the explicit rogue wave solutions of Eqs. (1) and (2),
the dynamics and some interesting structures of the rogue waves for Eqs. (1) and (2) from first to second order are
illustrated through some specific figures.

\section{Rogue wave solutions}
From the above section, we can observe that acquiring the adequate initial eigenfunction under the seed solutions enables us to obtain
rogue waves of Eqs. (1) and (2). To this end, we start from the periodic pane waves
\begin{equation}\label{25}
A[0]={\rm e}^{{\rm i}\theta} ,\ B[0]=-\frac{a}{\sqrt{1+a^{2}}},
\end{equation}
where $\theta=\displaystyle\frac{(a\sqrt{1+a^{2}}x+t)}{\sqrt{1+a^{2}}}$, and $a$ is a real constant. After that, inserting (\ref{25}) into the
Lax pair (\ref{04}) and (\ref{05}) and solving it, we have
\begin{equation}\label{26}
\Psi_{1}=\left(
\begin{array}{c}
(C_{1}{\rm e}^{N}-C_{2}{\rm e}^{-N}){\rm e}^{\frac{{\rm i}}{2}\theta} \\
(C_{1}{\rm e}^{-N}-C_{2}{\rm e}^{N}){\rm e}^{-\frac{{\rm i}}{2}\theta} \\
\end{array}
\right),
\end{equation}
where
$$
C_{1}=\frac{ (2\lambda+a-\sqrt{4\lambda^{2}+4a\lambda+1+a^{2}})^{\frac{1}{2}} }{\sqrt{4\lambda^{2}+4a\lambda+1+a^{2}}},\ C_{2}=\frac { (2\lambda+a+\sqrt{4\lambda^{2}+4a\lambda+1+a^{2}})^{\frac{1}{2}} }{\sqrt{4\lambda^{2}+4a\lambda+1+a^{2}}},
$$
and
$$
N=\frac{{\rm i}}{4\sqrt{1+a^{2}}\lambda}\sqrt{4\lambda^{2}+4a\lambda+1+a^{2}}(2\sqrt{1+a^{2}} \lambda x+t+\sum_{k=1}^{N}s_{k}f^{2k}).
$$
Here $f$ is a small real parameter, $s_{k}=m_{k}+{\rm i}n_{k},(m_{k},n_{k}\in\mathbb{R})$. Next, we fix $\lambda_{1}=-\displaystyle\frac{a}{2}+\displaystyle\frac{{\rm i}}{2}$, and set $\lambda=-\displaystyle\frac{a}{2}+\displaystyle\frac{{\rm i}}{2}+f^{2}$ in (\ref{26}). Then, the vector function $\Psi_{1}$ can be
expanded as Taylor series at $f=0$, that is
\begin{equation}\label{27}
\Psi_{1}(f)=\Psi_{1}^{[0]}+\Psi_{1}^{[1]}f^{2}+\Psi_{1}^{[2]}f^{4}+\cdots,
\end{equation}
here we firstly present the explicit expression of $\Psi_{1}^{[0]}$
$$
\psi_{1}^{[0]}=-\frac{\sqrt{2}}{2\sqrt{1+a^{2}}({\rm i}-a)}p_{1}^{[0]}{\rm e}^{\frac{{\rm i}}{2}\theta},\ \
\phi_{1}^{[0]}=\frac{\sqrt{2}}{2\sqrt{1+a^{2}}({\rm i}-a)}p_{2}^{[0]}{\rm e}^{-\frac{{\rm i}}{2}\theta},
$$
where
$$
\begin{array}{l}
p_{1}^{[0]}=(1-{\rm i})\sqrt{1+a^{2}}({\rm i}-a)x+(1-{\rm i})t+(1-{\rm i})\sqrt{1+a^{2}}({\rm i}-a),\\
p_{2}^{[0]}=(1-{\rm i})\sqrt{1+a^{2}}({\rm i}-a)x+(1-{\rm i})t-(1-{\rm i})\sqrt{1+a^{2}}({\rm i}-a).
\end{array}
$$
With the aid of the symbolic computation tool Maple, it is easy to verify that $\Psi_{1}^{[0]}=(\psi_{1}^{[0]},\phi_{1}^{[0]})$ is a nontrivial solution of
the Lax pair (\ref{04}) and (\ref{05}) with the seed solutions (\ref{25}) and the fixed spectral parameter $\lambda_{1}=-\displaystyle\frac{a}{2}+\displaystyle\frac{{\rm i}}{2}$. So that, by means of  the formulas (\ref{14}) and (\ref{15}),
we get
\begin{equation}\label{28}
A[1]={\rm e}^{{\rm i}\theta}(1+\frac{F_{1}+{\rm i}H_{1}}{D_{1}}),\ \ B[1]=\frac{1}{\sqrt{1+a^{2}}}\frac{G_{1}}{D_{1}^{2}},
\end{equation}
where
$$
\begin{array}{l}
F_{1}=(2a^{4}+4a^{2}+2)x^{2}-4\sqrt{1+a^2}axt+2t^{2}-2a^{4}-4a^{2}-2,\ \ H_{1}=4\sqrt{1+a^{2}}t,\\
D_{1}=-(a^{4}+2a^{2}+1)x^{2}+2a\sqrt{1+a^{2}}xt-t^{2}-a^{4}-2a^{2}-1,\\
G_{1}=-a(a^{2}+1)^{4}x^{4}+4a^{2}(a^{2}+1)^{5/2}tx^{3}-2a(a^{2}+1)((3a^{2}+1)t^{2}+a^{6}+5a^{4}+7a^{2}+3   )x^{2}+\\
~~~~~~4\sqrt{1+a^{2}}(a^{2}t^{2}+a^{6}+4a^{4}+5a^{2}+2)xt-at^{4}-(2a^{5}+8a^{3}+6a)t^{2}-a^{9}+6a^{5}+8a^{3}+3a,
\end{array}
$$
which is nothing but the first-order rogue wave solution of Eqs. (1) and (2), see Fig. 1, and it is direct to check that (\ref{28})
satisfies Eqs. (1), (2) and (3).

Next, in order to obtain the second-order rogue wave solution of Eqs. (1) and (2), $\Psi_{1}^{[1]}$ should be used to construct the generating function,
$$
\psi_{1}^{[1]}=\displaystyle\frac{\sqrt{2}(1+{\rm i})}{12({\rm i}-a)^{3}(1+a^{2})^{3/2}}p_{1}^{[1]}{\rm e}^{\frac{{\rm i}}{2}\theta},\ \
\phi_{1}^{[1]}=-\displaystyle\frac{\sqrt{2}(1+{\rm i})}{12({\rm i}-a)^{3}(1+a^{2})^{3/2}}p_{2}^{[1]}{\rm e}^{-\frac{{\rm i}}{2}\theta},
$$
where
$$\begin{array}{l}
p_{1}^{[1]}=\sqrt{1+a^{2}}({\rm i}-a)[(a^{4}-2{\rm i}a^{3}-2{\rm i}a-1)x^{3}+(3a^{4}-6{\rm i}a^{3}-6{\rm i}a-3)x^{2}+(3a^{4}-6{\rm i}a^{3}-6{\rm i}a-3)x
+(3x+3)t^{2}\\~~~~~~-3a^{4}+6{\rm i}a^{3}+6{\rm i}a+3]+t^{3}+((3a^{4}-6{\rm i}a^{3}-6{\rm i}a-3)x^{2}+(6a^{4}-12{\rm i}a^{3}-12{\rm i}a-6)x+3a^{4}+6{\rm i}a^{3}+12a^{2}\\~~~~~~+6{\rm i}a+9)t+6{\rm i}m_{1}a^{4}-6n_{1}a^{4}+12{\rm i}n_{1}a^{3}+12m_{1}a^{3}+12{\rm i}n_{1}a+12m_{1}a-6{\rm i}m_{1}+6n_{1}  ,\\
p_{2}^{[1]}=\sqrt{1+a^{2}}({\rm i}-a)[(a^{4}-2{\rm i}a^{3}-2{\rm i}a-1)x^{3}-(3a^{4}-6{\rm i}a^{3}-6{\rm i}a-3)x^{2}+(3a^{4}-6{\rm i}a^{3}-6{\rm i}a-3)x
+(3x-3)t^{2}\\~~~~~~+3a^{4}-6{\rm i}a^{3}-6{\rm i}a-3]+t^{3}+((3a^{4}-6{\rm i}a^{3}-6{\rm i}a-3)x^{2}-(6a^{4}-12{\rm i}a^{3}-12{\rm i}a-6)x+3a^{4}+6{\rm i}a^{3}+12a^{2}\\~~~~~~+6{\rm i}a+9)t+6{\rm i}m_{1}a^{4}-6n_{1}a^{4}+12{\rm i}n_{1}a^{3}+12m_{1}a^{3}+12{\rm i}n_{1}a+12m_{1}a-6{\rm i}m_{1}+6n_{1}.
\end{array}
$$
By using  the following limit process
$$\begin{array}{l}
\lim\limits_{f\rightarrow0}\frac{\displaystyle T[1]|_{\lambda=-a/2+{\rm i}/2+f^{2}}\Psi_{1}}{\displaystyle f^{2}}=\lim\limits_{f\rightarrow0}
\frac{\displaystyle(f^{2}+T_{1}[1])\Psi_{1}}{\displaystyle f^{2}}=
\Psi_{1}^{[0]}+T_{1}[1]\Psi_{1}^{[1]}\equiv\Psi_{1}[1],
\end{array}$$
we have   
\begin{equation}\label{29}
\psi_{1}[1]=\frac{\sqrt{2}(-1+{\rm i})}{6({\rm i}-a)^{2}(1+a^{2})^{3/2}D_{1}}\rho_{1}{\rm e}^{\frac{{\rm i}}{2}\theta},\ \
\phi_{1}[1]=\frac{\sqrt{2}(-1+{\rm i})}{6({\rm i}-a)^{2}(1+a^{2})^{3/2}D_{1}}\rho_{2}{\rm e}^{-\frac{{\rm i}}{2}\theta},
\end{equation}
where
$$
\begin{array}{l}
\rho_{1}=\sqrt{1+a^{2}}[(-a^{8}+2{\rm i}a^{7}-2a^{6}+6{\rm i}a^{5}+6{\rm i}a^{3}+2a^{2}+2{\rm i}a+1)x^{4}+(-2a^{8}+4{\rm i}a^{7}-4a^{6}+12{\rm i}a^{5}+12{\rm i}a^{3}+4a^{2}\\~~~~~+4{\rm i}a+2)x^{3}+(-6a^{8}+12{\rm i}a^{7}-12a^{6}+36{\rm i}a^{5}+36{\rm i}a^{3}+12a^{2}+12{\rm i}a+6)x-t^{4}+
((-6a^{4}+6{\rm i}a^{3}-6a^{2}+6{\rm i}a)x^{2}\\~~~~~+(-6a^{4}+12{\rm i}a^{3}+12{\rm i}a+6)x+6{\rm i}a^{3}+6a^{2}+6{\rm i}a+6)t^{2}+(3{\rm i}m_{1}a^{4}-3n_{1}a^{4}+6m_{1}a^{3}+6{\rm i}n_{1}a^{3}+6{\rm i}n_{1}a+6m_{1}a\\~~~~~-3{\rm i}m_{1}+3n_{1})t-3a^{8}+6{\rm i}a^{7}-6a^{6}+18{\rm i}a^{5}+18{\rm i}a^{3}+6a^{2}+6{\rm i}a+3]+
(3n_{1}a^{7}-3{\rm i}m_{1}a^{7}-3m_{1}a^{6}-3{\rm i}n_{1}a^{6}\\~~~~~+9n_{1}a^{5}-9{\rm i}m_{1}a^{5}-9m_{1}a^{4}-9{\rm i}n_{1}a^{4}-9{\rm i}m_{1}a^{3}
+9n_{1}a^{3}-9{\rm i}n_{1}a^{2}-9m_{1}a^{2}+3n_{1}a-3{\rm i}m_{1}a-3m_{1}-3{\rm i}n_{1})x\\~~~~~+
((4a^{3}-2{\rm i}a^{2}+4a-2{\rm i})x+2a^{3}-4{\rm i}a^{2}+2a-4{\rm i})t^{3}+
((4a^{7}-6{\rm i}a^{6}+8a^{5}-14{\rm i}a^{4}+4a^{3}-10{\rm i}a^{2}-2{\rm i})x^{3}\\~~~~~+(6a^{7}-12{\rm i}a^{6}+6a^{5}-24{\rm i}a^{4}-6a^{3}-12{\rm i}a^{2}-6a)x^{2}
-(6{\rm i}a^{6}+18{\rm i}a^{4}+18{\rm i}a^{2}+6{\rm i})x+6a^{7}+18a^{5}+18a^{3}\\~~~~~+6a)t+
3{\rm i}m_{1}a^{7}-3n_{1}a^{7}+3m_{1}a^{6}+3{\rm i}n_{1}a^{6}-9n_{1}a^{5}+9{\rm i}m_{1}a^{5}+9m_{1}a^{4}+9{\rm i}n_{1}a^{4}+9{\rm i}m_{1}a^{3}-9n_{1}a^{3}+9{\rm i}n_{1}a^{2}\\~~~~~+9m_{1}a^{2}-3n_{1}a
+3{\rm i}m_{1}a+3m_{1}+3{\rm i}n_{1},\\
\rho_{2}=\sqrt{1+a^{2}}[(-a^{8}+2{\rm i}a^{7}-2a^{6}+6{\rm i}a^{5}+6{\rm i}a^{3}+2a^{2}+2{\rm i}a+1)x^{4}+(2a^{8}-4{\rm i}a^{7}+4a^{6}-12{\rm i}a^{5}-12{\rm i}a^{3}-4a^{2}\\~~~~~-4{\rm i}a-2)x^{3}+(6a^{8}-12{\rm i}a^{7}+12a^{6}-36{\rm i}a^{5}-36{\rm i}a^{3}-12a^{2}-12{\rm i}a-6)x-t^{4}
+((-6a^{4}+6{\rm i}a^{3}-6a^{2}+6{\rm i}a)x^{2}\\~~~~~+(6a^{4}-12{\rm i}a^{3}-12{\rm i}a-6)x+6{\rm i}a^{3}+6a^{2}+6{\rm i}a+6)t^{2}+
(3{\rm i}m_{1}a^{4}-3n_{1}a^{4}+6{\rm i}n_{1}a^{3}+6m_{1}a^{3}+6m_{1}a+6{\rm i}n_{1}a\\~~~~~+3n_{1}-3{\rm i}m_{1})t
-3a^{8}+6{\rm i}a^{7}-6a^{6}+18{\rm i}a^{5}+18{\rm i}a^{3}+6a^{2}+6{\rm i}a+3]+
(-3{\rm i}m_{1}a^{7}+3n_{1}a^{7}-3{\rm i}n_{1}a^{6}-3m_{1}a^{6}\\~~~~~-9{\rm i}m_{1}a^{5}+9n_{1}a^{5}-9{\rm i}n_{1}a^{4}-9m_{1}a^{4}-9{\rm i}m_{1}a^{3}+9n_{1}a^{3}-9m_{1}a^{2}-9{\rm i}n_{1}a^{2}-3{\rm i}m_{1}a+3n_{1}a-3m_{1}-3{\rm i}n_{1})x\\~~~~~+
((4a^{3}-2{\rm i}a^{2}+4a-2{\rm i})x-2a^{3}+4{\rm i}a^{2}-2a+4{\rm i})t^{3}+
((4a^{7}-6{\rm i}a^{6}+8a^{5}-14{\rm i}a^{4}+4a^{3}-10{\rm i}a^{2}-2{\rm i})x^{3}\\~~~~~+(-6a^{7}+12{\rm i}a^{6}-6a^{5}+24{\rm i}a^{4}+6a^{3}+12{\rm i}a^{2}+6a)x^{2}+
(-6{\rm i}a^{6}-18{\rm i}a^{4}-18{\rm i}a^{2}-6{\rm i})x-6a^{7}-18a^{5}-18a^{3}\\~~~~~-6a)t
-3{\rm i}m_{1}a^{7}+3n_{1}a^{7}-3m_{1}a^{6}-3{\rm i}n_{1}a^{6}+9n_{1}a^{5}-9{\rm i}m_{1}a^{5}-9m_{1}a^{4}-9{\rm i}n_{1}a^{4}
-9{\rm i}m_{1}a^{3}+9n_{1}a^{3}-9{\rm i}n_{1}a^{2}\\~~~~~-9m_{1}a^{2}+3n_{1}a-3{\rm i}m_{1}a-3m_{1}-3{\rm i}n_{1},
\end{array}
$$
then  a special solution of the Lax pair (\ref{04}) and (\ref{05}) with $A[1]$, $B[1]$ and $\lambda=\lambda_{1}=-a/2+{\rm i}/2$ can be obtained.
Thus,  the second-order rogue wave solution of Eqs. (1) and (2) can be given by substituting (\ref{29}) and (\ref{28}) into (\ref{17}) and (\ref{18}).
Here we omit the explicit expressions of $A[2]$ and  $B[2]$  because it is rather tedious and inconvenient to write them down  here, but
it is not difficult to verify that they satisfy Eqs. (1), (2) and (3) with the help of Maple. Finally, we show some interesting structures of the second-order
rogue waves, see Figs. 2-4. For the second-order rogue wave solution of Eqs. (1) and (2) under the normalization condition (\ref{03}), there are
two free parameters in the expressions of $A[2]$ and $B[2]$ except $a$, namely, $m_{1}$ and $n_{1}$.  When we set $m_{1}=0,~n_{1}=0,$ the fundamental
second-order rogue waves can be given, see Fig. 2. When the parameter are chosen by $m_{1}\neq0,~n_{1}=0$, the second-order rogue waves of triangular patterns
can be presented, see Fig. 3-4.

\section{Conclusion}
In this paper, the AB system which is important
to describe marginally unstable baroclinic wave packets in geophysical fluids, ultra-short optical pulse propagation in nonlinear optics,
and also mesoscale gravity current transmission on a sloping bottom in the problem of cold gravity current
is investigated through the so-called generalized DT. We find a unified formula to construct $N$th-order rogue wave solution for Eqs. (1) and (2) by the direct iterative rule. As application, rogue wave solutions from first to second order are obtained.
With the help of some  free parameters, the first-order rogue waves of fundamental pattern, the second-order rogue waves of
fundamental and triangular patterns are shown, respectively.
The results further reveal and enrich the dynamical properties of Eqs. (1) and (2), and we hope our results will be verified in real experiments in the future.
Besides, on the one hand, continuing the generalized DT one by one, the higher-order rogue waves of Eqs. (1) and (2) can be generated,
and they are likely to possess the more abundant dynamic properties, such as the \lq\lq claw\rq\rq,  \lq\lq claw-line\rq\rq, and  \lq\lq claw-arc\rq\rq structures like the high-order rogue waves of the standard NLS equation \cite{35}. On the other hand,
motivated by the remarkable work of Baronio and Guo et al. for the Manakov equations \cite{29,30}, the interactions between the rogue waves and the
solitons or the breathers of Eqs. (1) and (2) may also be obtained by the darboux transformation.
Both of these problems are interesting, and we will investigate them in our future papers.

\section*{Acknowledgment}
The project is supported by the National Natural Science Foundation of China (Grant Nos. 11075055,
11275072), Innovative Research Team Program of the National Science Foundation of China (No.
61021104), National High Technology Research and Development Program (No. 2011AA010101),
Shanghai Knowledge Service Platform for Trustworthy Internet of Things (No. ZF1213), Talent Fund
and K.C. Wong Magna Fund in Ningbo University.

\begin{figure}
\centering
\renewcommand{\figurename}{{\bf Fig.}}
{\includegraphics[height=3cm,width=4cm]{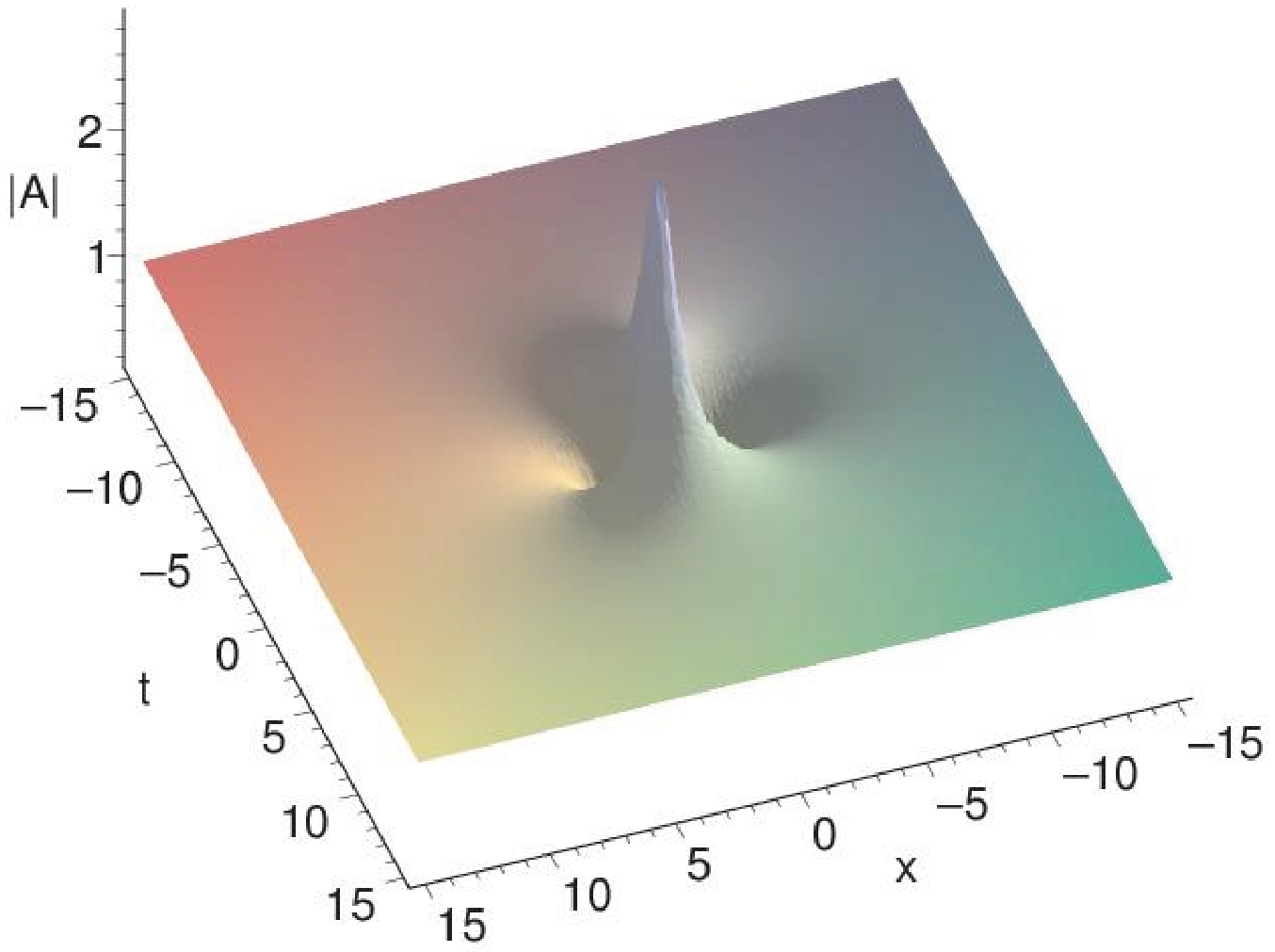}}~~~~~~~~~~~~~~~~~~~~~~~~
{\includegraphics[height=3cm,width=4cm]{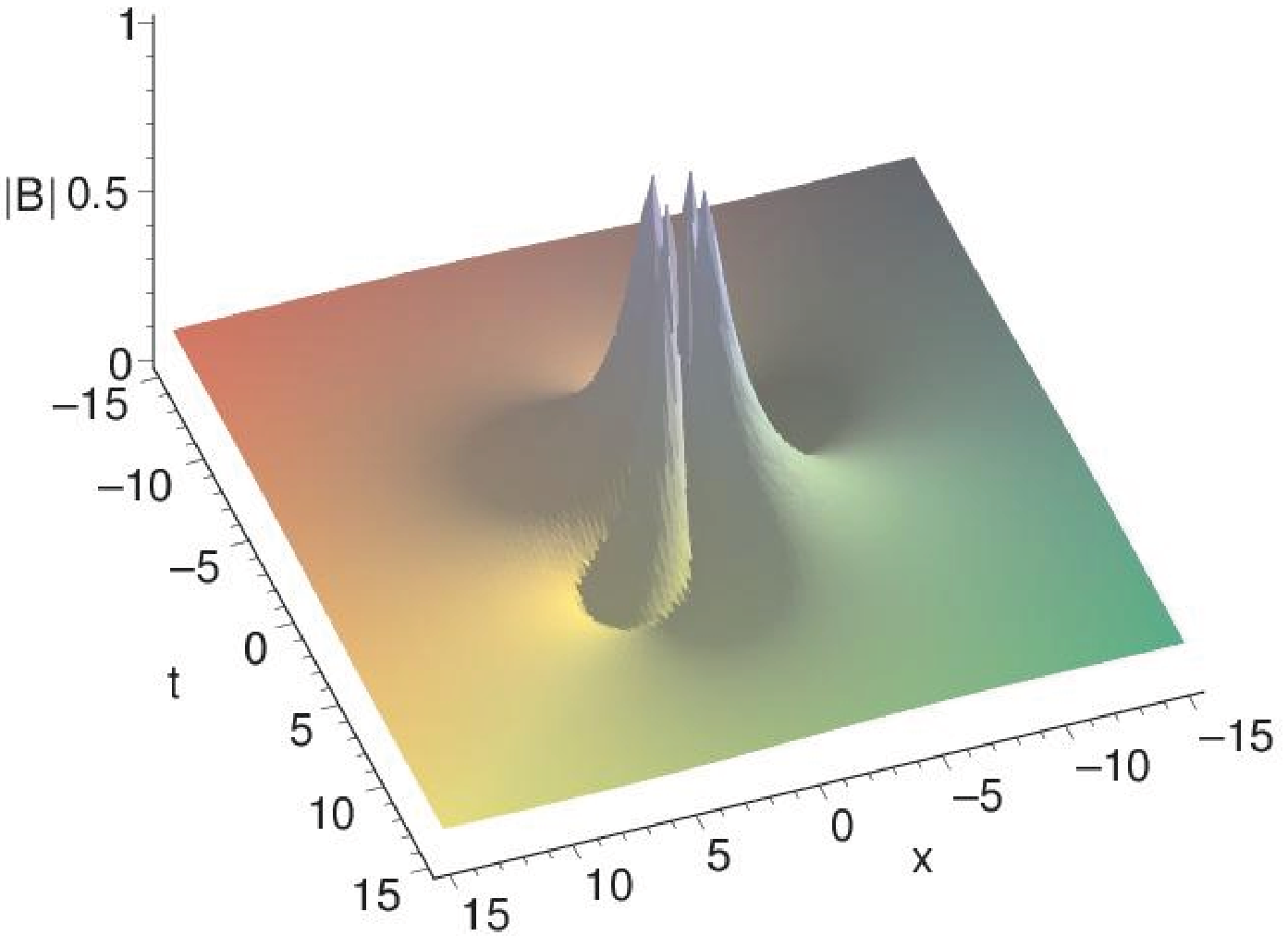}}
\begin{center}
\hskip 1cm $(\rm{a})$ \hskip 6cm $(\rm{b})$
\end{center}
\caption{The first-order rogue waves of the AB system. (\rm{a}) Rogue wave in $A$ component; (\rm{b})  Rogue wave in $B$ component. The parameters are  $a=1/10$.}
\centering
\renewcommand{\figurename}{{\bf Fig.}}
{\includegraphics[height=3cm,width=4cm]{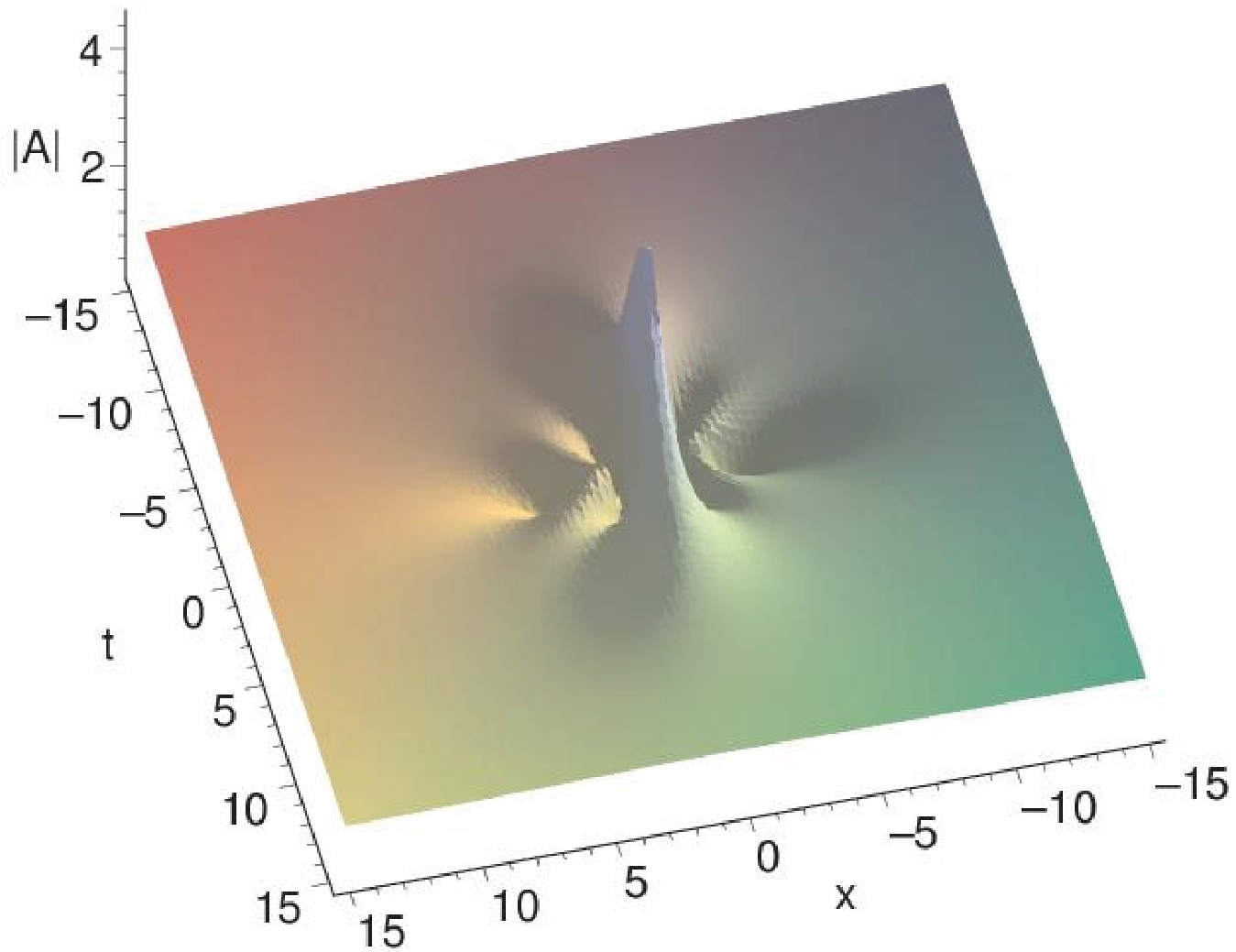}}~~~~~~~~~~~~~~~~~~~~~~~~
{\includegraphics[height=3cm,width=4cm]{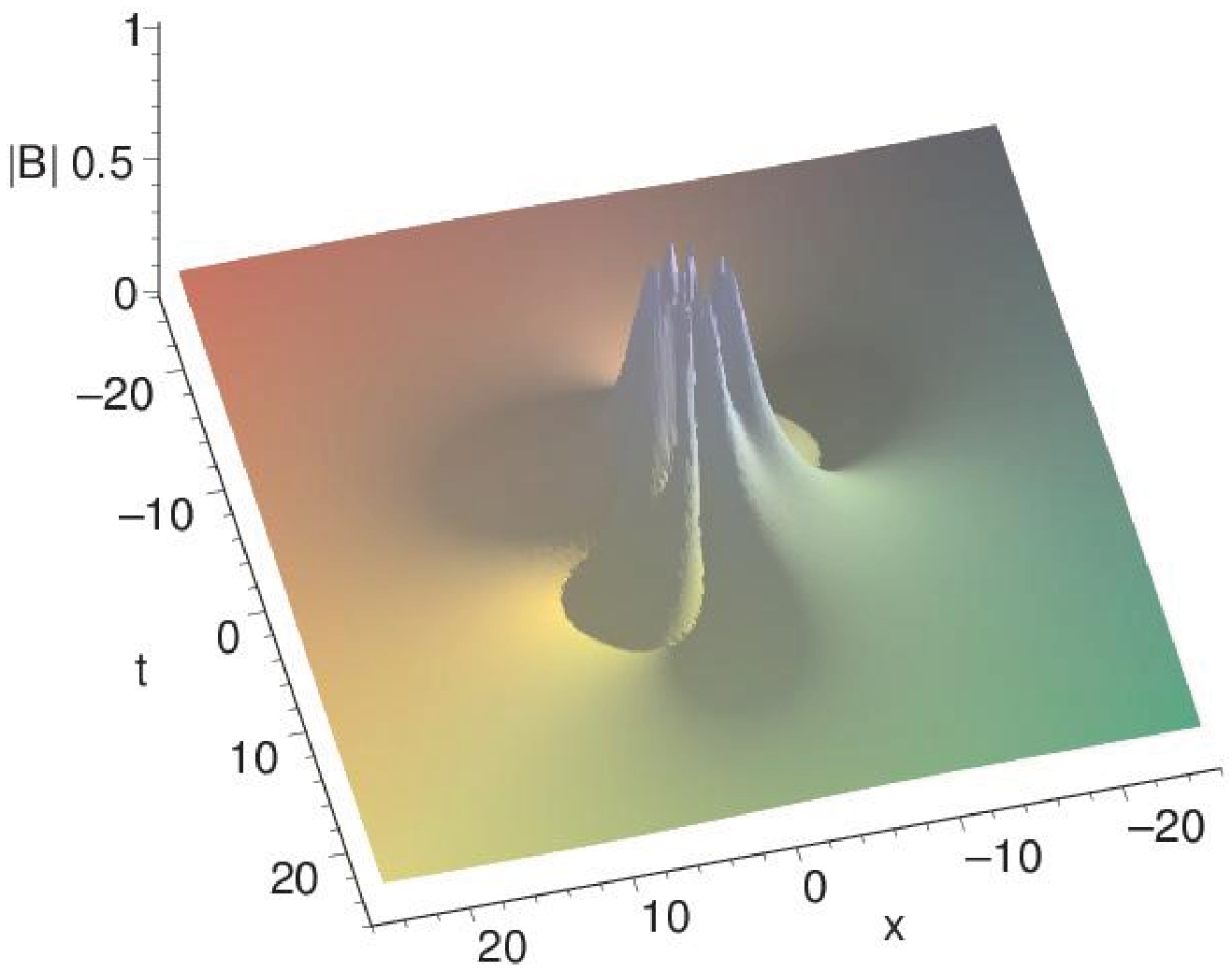}}
\begin{center}
\hskip 1cm $(\rm{a})$ \hskip 5cm $(\rm{b})$
\end{center}
\caption{The second-order rogue waves of the AB system. (\rm{a}) Rogue wave in $A$ component; (\rm{b})  Rogue wave in $B$ component. The parameters are  $a=1/10,~m_{1}=0,~n_{1}=0$. }
\centering
\renewcommand{\figurename}{{\bf Fig.}}
{\includegraphics[height=3cm,width=4cm]{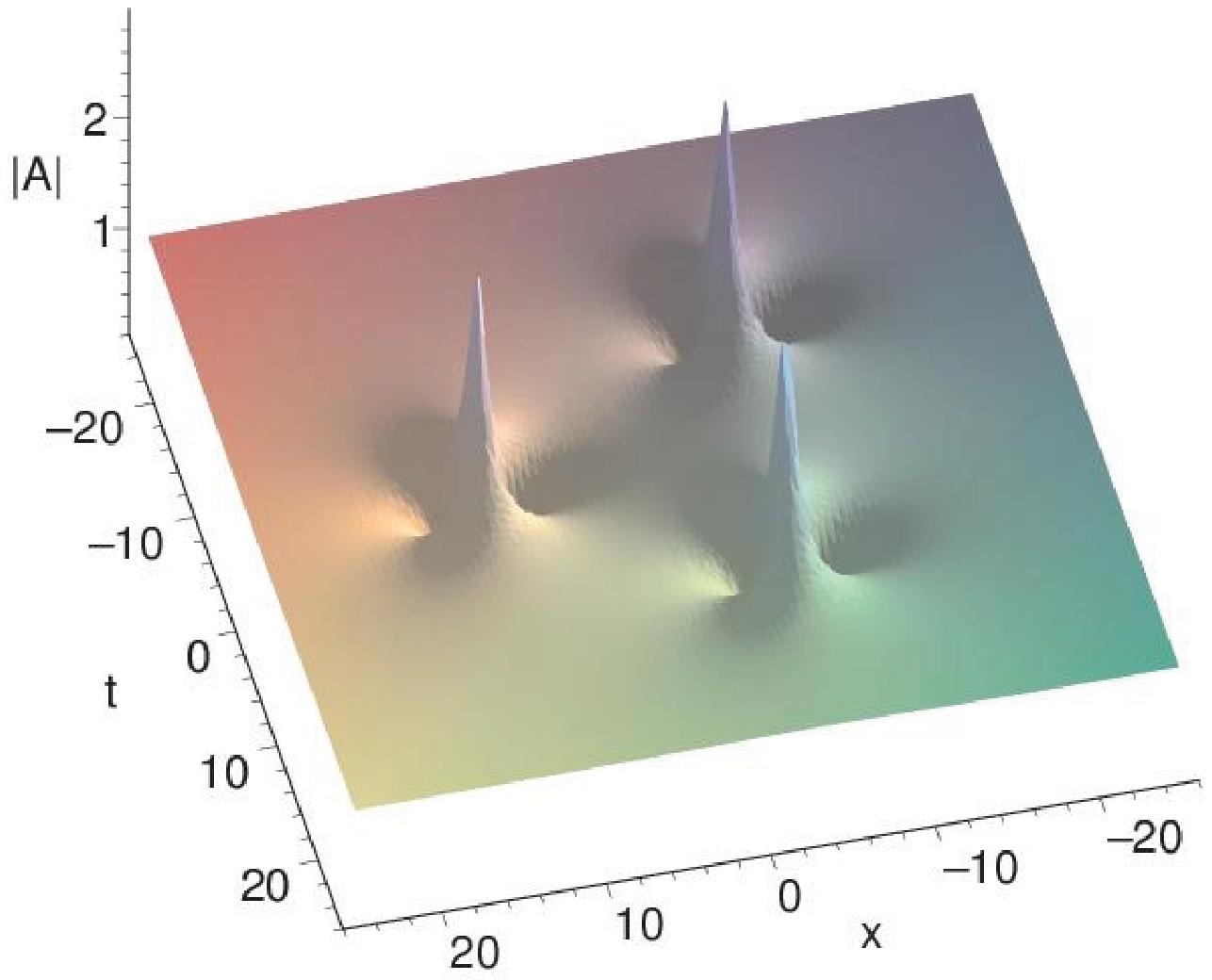}}~~~~~~~~~~~~~~~~~~~~~~~~
{\includegraphics[height=3cm,width=4cm]{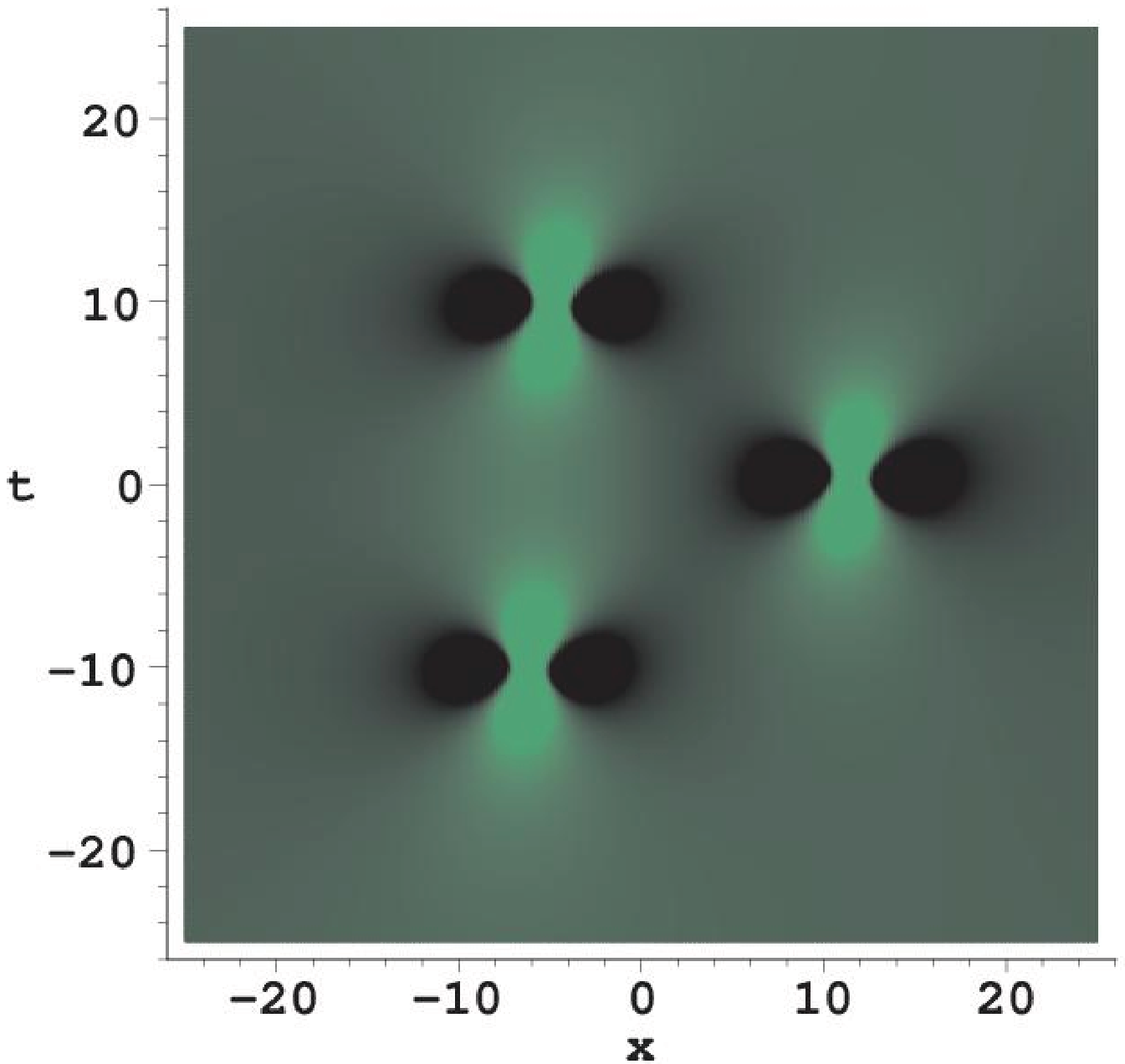}}
\begin{center}
\hskip 1cm $(\rm{a})$ \hskip 6cm $(\rm{b})$
\end{center}
\caption{The second-order rogue waves of triangular pattern for the AB system in $A$ component; The parameters are  $a=1/10,~m_{1}=500,~n_{1}=0$.}
\centering
\renewcommand{\figurename}{{\bf Fig.}}
{\includegraphics[height=3cm,width=4cm]{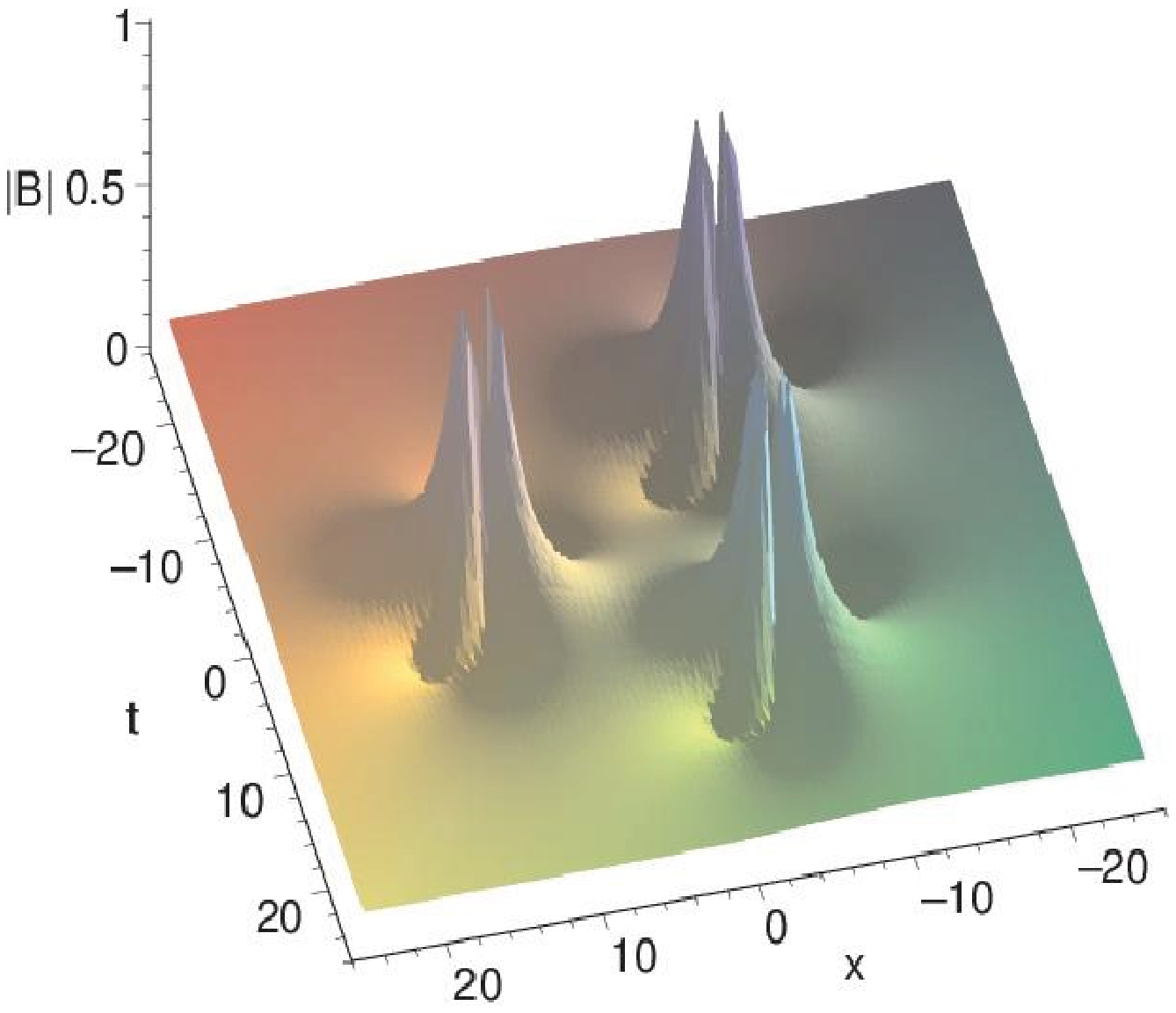}}~~~~~~~~~~~~~~~~~~~~~~~~
{\includegraphics[height=3cm,width=4cm]{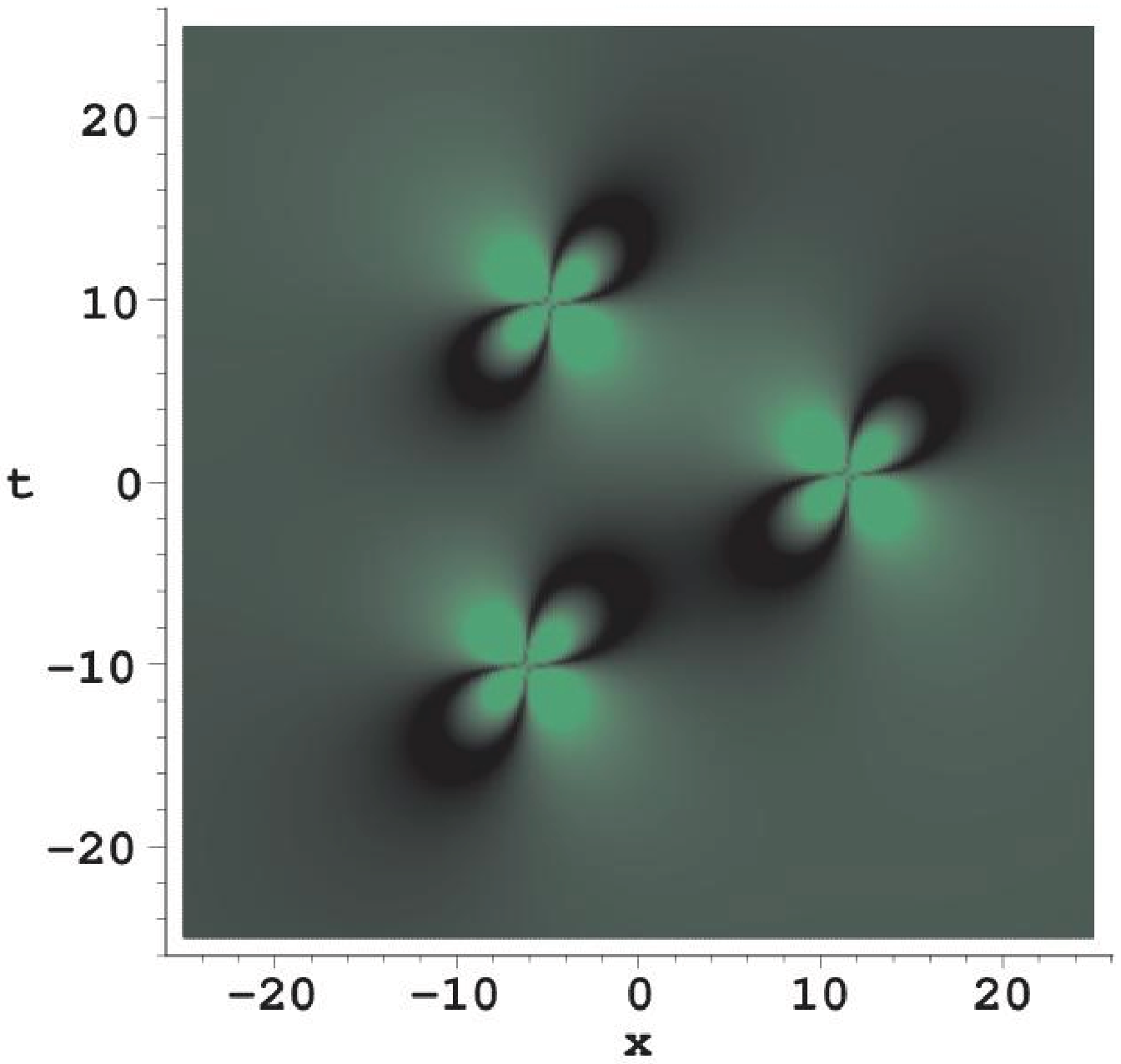}}
\begin{center}
\hskip 1cm $(\rm{a})$ \hskip 6cm $(\rm{b})$
\end{center}
\caption{The second-order rogue waves of triangular pattern for the AB system in $B$ component; The parameters are  $a=1/10,~m_{1}=500,~n_{1}=0$.}
\end{figure}

\end{document}